# Multi-wavelength high energy gas-filled fiber Raman laser spanning from 1.53 µm to 2.4 µm

ABUBAKAR I. ADAMU,[1,*] YAZHOU WANG,[1] MD. SELIM HABIB,[2] MANOJ. K. DASA,[1] J. E ANTONIO-LOPEZ,[3] RODRIGO AMEZCUA-CORREA,[3] OLE BANG,[1,4] AND CHRISTOS MARKOS[1,4, *]

[1]*DTU Fotonik, Technical University of Denmark, DK-2800 Kgs. Lyngby, Denmark*
[2]*Department of Electrical and Computer Engineering, Florida Polytechnic University, Lakeland, FL-33805, USA*
[3]*CREOL, The College of Optics and Photonics, University of Central Florida, Orlando, FL-32816, USA*
[4]*NORBLIS IVS, Virumgade 35D, DK-2830 Virum, Denmark.*
*Corresponding author: abisa@fotonik.dtu.dk, chmar@fotonik.dtu.dk*

**In this work, we present a high pulse energy multi-wavelength Raman laser spanning from 1.53 µm up to 2.4 µm by employing the cascaded rotational stimulated Raman scattering (SRS) effect in a 5-m hydrogen ($H_2$) - filled nested anti-resonant fiber (NARF), pumped by a linearly polarized Er/Yb fiber laser with a peak power of ~13 kW and pulse duration of ~7 ns in the C-band. The developed Raman laser has distinct lines at 1683 nm, 1868 nm, 2100 nm, and 2400 nm, with pulse energies as high as 18.25 µJ, 14.4 µJ, 14.1 µJ, and 8.2 µJ, respectively. We demonstrate how the energy in the Raman lines can be controlled by tuning the $H_2$ pressure from 1 bar to 20 bar.** ©2020

https://doi.org

Gas-filled hollow-core fiber lasers have been demonstrated to be a vital substitute to free-space lasers and solid-core fiber lases [1]. They possess intrinsic advantages over their counterparts, such as high damage threshold, efficient heat dispassion capacity through gas circulation, tunable nonlinear and dispersion properties, and potential of changing wavelength coverage by changing from active or inactive gasses [2]. By confining light within a micrometer-sized fiber core (air) region, a significant and enhanced light-matter interaction over extensive length can be achieved, compared to the conventional free-space structure, therefore enabling efficient SRS processes. This type of Raman lasers was initially reported in the short wavelength region (~ 500 nm to ~ 1.2 µm) [3–6], and subsequently extended towards longer wavelengths [7–16]. A common wavelength of this kind of laser is at 1.9 µm [8–12], which is directly and efficiently generated by pumping at 1.06 µm with a fiber laser and using $H_2$ as the active gas medium, which has the highest Stokes shift coefficient (4155 cm$^{-1}$) of any known gas. Moving towards longer wavelength above 2.0 µm was a long-standing challenge due to the high silica loss in this spectral region. This limitation was addressed by the recent advent of negative curvature anti-resonant (AR) hollow-core fibers (HCFs), such as NARF, which were able to effectively suppress the high silica loss by using a properly designed anti-resonant structure [17,18]. A few reports on mid-infrared fiber Raman lasers have already been reported [19–22], based on the vibrational SRS process in $H_2$ or methane ($CH_4$). In these reports, the quantum efficiency could easily reach a desirable level (e.g., >60%), leading to high pulse energies in the micro-joule range. However, their wavelengths are mainly located at two regions i.e., 2.8 µm and 4.2 µm - 4.4 µm, as a result of limitations on pump wavelength selection combined with the fixed shift of the Stokes coefficient of the used gases.

On the other hand, the efficient SRS process in gas-filled HCFs can easily lead to the generation of cascaded Raman Stokes lines [4,5,21,23], consequently constituting a promising way for multi-wavelength (or frequency comb-like) laser emission, which has versatile applications such as sensing, wavelength division multiplexing communication, and optical signal processing [24]. Moreover, when the phase-matching condition of anti-Stokes line generation is satisfied by controlling the gas-pressure dependent dispersion and utilizing the dispersion difference of different order modes, the generated Stokes lines can also lead to the formation of multiple anti-Stokes lines through four-wave mixing involving two pump photons, a Stokes photon and an anti-Stokes photon [14,25,26]. In comparison with the conventional method of selecting multiple wavelengths with a comb filter within the limited gain bandwidth of a rare-earth ion-doped gain medium [27,28], this method can achieve multi-wavelength operation in a much broader wavelength range. Despite this, the reported energy/intensity of high-order Stokes and anti-Stokes lines were quite low when compared to the first Stokes line [14,25,26]. Besides, the generated wavelengths of all these multi-wavelength cascaded based Raman lines reported were limited to less than ~2 µm. However, advances in AR-HCF has led to fibers with improved transmission in the near to mid- IR region. Exceptional examples are the results in Refs. [21], and [22]. In [21], a second-order Stokes line at 2.8 µm was generated from a first-order Stokes line at 1.5 µm, which in turn was generated from a high-power pump laser at 1.06 µm. However, only two wavelengths were achieved in this demonstration, since the third-order Stokes line goes up to ~15 µm wavelength where the silica HCF has a high loss. In Ref. [22], three Raman lines at 2.9 µm, 3.3 µm, and 3.5 µm were achieved by cross-utilizing the SRS effect of $H_2$ and deuterium, however, their pulse energies were in a relatively low level of 0.84 µJ, 0.34 µJ and 1.28 µJ respectively. According to the literature [4,5,14,22,25,26], demonstrating an efficient and high pulse energy multi-wavelength gas-filled Raman laser in a broad wavelength range would be of significant importance to the laser community by providing high energy laser pulses at unique wavelengths. In this letter, we demonstrate a cascaded Raman laser emitting at multiple wavelengths of 1.68 µm, 1.87 µm, 2.1 µm, and 2.4 µm. The quantum efficiencies are as high as 50.5%, 43.5%, 52.5%, and 38.1%, respectively, corresponding to high pulse energies of 18.5 µJ, 14.4 µJ, 14.1 µJ and 8.3 µJ, respectively.

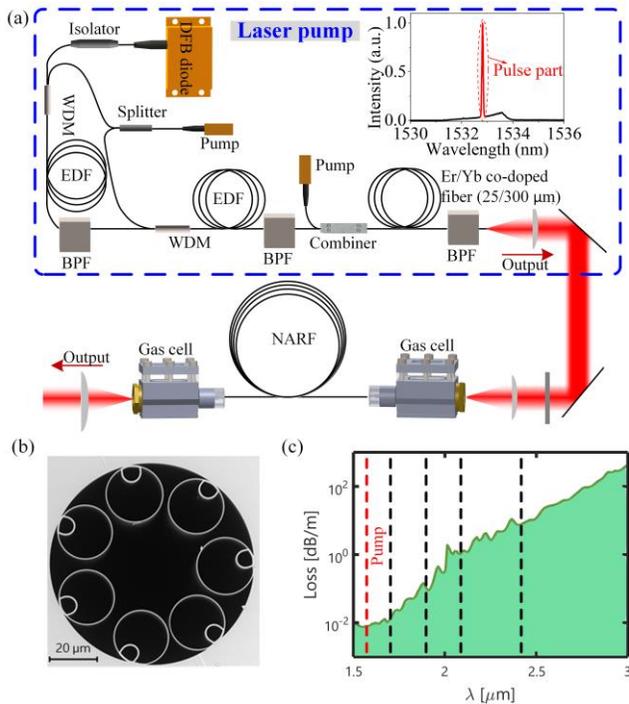

Figure 1. (a) Experimental setup with pump laser (blue box, inset shows the output spectrum), followed by the NARF-based gas cell for Raman generation. (b) Scanning electron microscopy (SEM) image of the NARF with core diameter 37 µm and capillary wall thicknesses of ~406 nm and ~630 nm for the large capillary and nested capillary, respectively (scale bar: 20 µm). (c) Simulated attenuation spectrum of the NARF. The red dashed line is the pump wavelength and the black dashed lines indicate the wavelengths of the cascaded Stokes lines.

The 1532.8 nm pump laser shown in Fig. 1(a) is custom-built using a directly modulated DFB diode-based, all-polarization maintaining master oscillator power amplifier (MOPA) configuration. It is operated by a seed oscillator that delivers linearly polarized pulses at 8 kHz repetition rate and 6.9 ns pulse duration. The seed laser is first pre-amplified with two stages of Er-doped fiber (EDF) pre-amplifiers core-pumped by sharing the same CW pump laser diode through a fiber splitter, and then amplified by an Erbium/Ytterbium (Er/Yb) co-doped double-clad fiber power amplifier cladding-pumped by a 10 W CW laser diode at 915 nm. The final average output power of the fiber laser is 1.4 W. By excluding the contribution of ASE, the average power of the pure pulse signal part is estimated to be ~640 mW, corresponding to ~13 kW peak power and ~80 µJ pulse energy. The linewidth measured by an optical spectrum analyzer (ANDO AQ6317B, AssetRelay) with a resolution of 0.01 nm, is ~0.06 nm.

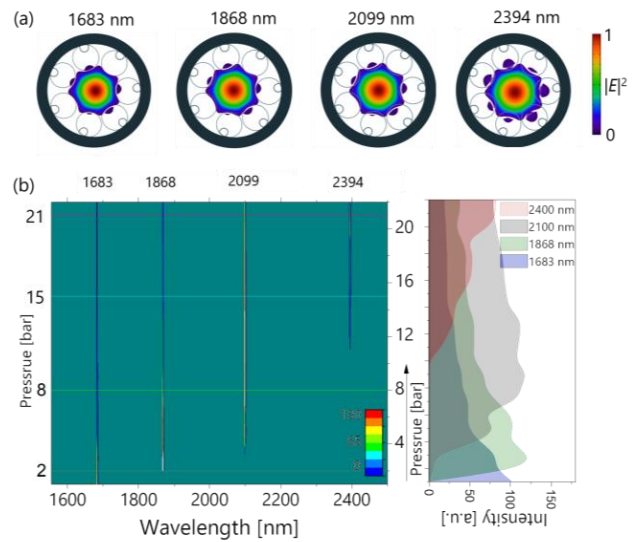

Figure 2. (a) Simulation of the fundamental mode profiles at central wavelengths of each Raman Stokes line. (b) Experimental result showing the evolution of the Raman laser spectrum as a function of $H_2$ pressure. Horizontal lines at 2, 8, 15 and 21 bar represent the graphs in Fig. 3. The right graph depicts the corresponding evolution of each Stokes line intensity with pressure.

In the experiment, the pump laser is focused into a 5 m NARF (SEM image of the cross-section shown in Fig. 1(b)) with a coupling efficiency of ~80%. Such a fiber length, combined with the relatively small core size of ~37 µm, is experimentally found to be sufficient to generate several high order Raman Stokes lines. The fiber has a loss of ~ 8 dB/km at the pump wavelength as shown from simulation result in Fig. 1(c), which increases gradually towards longer wavelengths. The nested feature of the fiber capillaries provides not only enhanced transmission, but also a lower bending loss [29], which allows the fiber to be coiled with a 35 cm radius for the experiments. The mode profiles at several Stokes wavelengths, shown in Fig. 2(a), are simulated based on finite element method using commercial software [20]. A quarter-wave plate is placed before the NARF to convert the pump light from linear polarization to circular polarization and optimize the polarization orientation, to enhance the rotational SRS effect.

The evolution of the measured Raman spectrum in terms of pressure was investigated at the maximum pump power and proper orientation of the quarter-wave plate, as presented in Fig. 2. An optical spectrum analyzer (OSA) (Spectro 320, Instrument Systems) was used to record the spectra. Initially, only the first-order rotational Stokes line at 1683 nm was observed at the pressure of ~1 bar. Increasing the gas pressure increases the Raman gain coefficient of the steady-state Raman regime [30] and suppresses the transient Raman regime [19], which has a lower gain coefficient than the steady-state regime. As a result, the intensity of the first-order Stokes line first continuously increases until it is sufficient to serve as a new pump to generate the second-order Stokes line at 1868 nm. Similarly, the third-order Stokes line at 2099 nm was formed from the second-order Stokes line and then increases in intensity and finally leads to the formation of a fourth-order Stokes line at 2394 nm when the pressure increases to ~11 bar. It can be seen that the emergence of new Stokes lines always is accompanied by a decrease in the intensity of the previous order Stokes line, indicating that the last order Stokes plays the role of the pump for the generation of the new Stokes line. Because the fifth-order Stokes line at ~2.8 μm could not be formed due the high fiber loss of 105 dB/m, the intensity of the fourth-order Stokes line continuously increases with pressure up to ~21 bar, which is the limit of the experimental setup.

For better clarity, Fig. 3 shows the typical spectra at different pressures corresponding to the horizontal lines in the contour plot in Fig. 2(b). The variation of the relative intensity of these laser lines implies that the pulse energy of each laser line could be accurately controlled by changing the $H_2$ pressure. Note that the FWHM of each Stokes line is in the level of MHz [31], which is less than the resolution of the OSA and therefore couldn't be measured here. Figure 4(a) shows a typical pulse profile of the first-order Stokes line at the $H_2$ pressure of ~1 bar, as well as the corresponding pulse profile of the original pump before the NARF, measured by a 1.2 GHz near-IR photodetector (DET01CFC, Thorlabs), connected to a 4 GHz oscilloscope (Teledyne Lecroy HD09000). This Stokes pulse exhibits a single peak profile, with a pulse duration of 4.9 ns, which is obviously less than the original pump pulse duration of ~7 ns. Besides the most frequently observed Gaussian-like pulse profile, some occasional Raman pulses with a different shape were also detected, such as the pulse profiles shown in Fig. 4(b) and (c), where they are characterized by typical multi-peak structure. This originates from the random nature of the quantum noise [32], which initiates the SRS process. Although the measured Raman pulse profiles have various shapes, their pulse duration always appears less than the pump laser, since the Stokes pulse is formed in the central part of the pump pulse where the peak power is sufficient to enable the SRS process. For the same reason, the high order Stokes line also has a narrower pulse duration than its pump, i.e., the preceding order Stokes line. Therefore, the 4th-order Stokes here is anticipated to have the narrowest pulse duration when compared to 1-3rd Stokes. This kind of reduction in pulse duration is another limitation on the generation of higher-order Stokes lines, since the narrower pulse duration is accompanied by a higher contribution of the transient Raman regime.

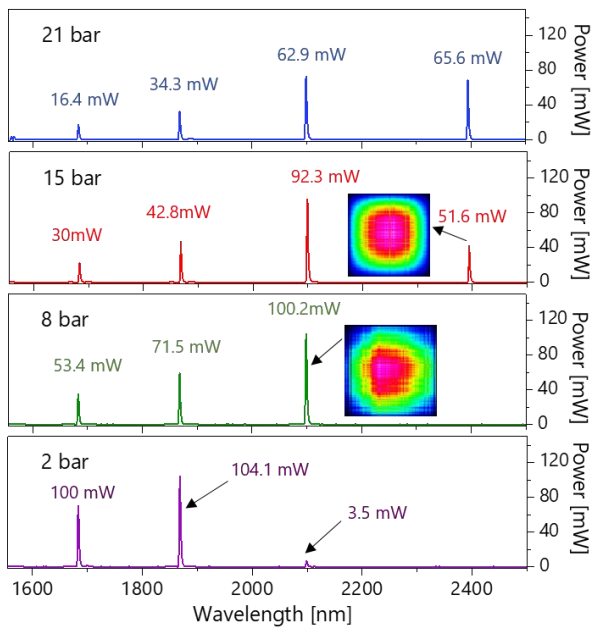

Figure 3. Optical spectra of the cascaded Raman fiber laser at different pressures. The spectra are taken at the pressures marked by horizontal lines in Fig. 2(b). Mode profiles of 2100 nm and 2400 nm Raman laser are shown.

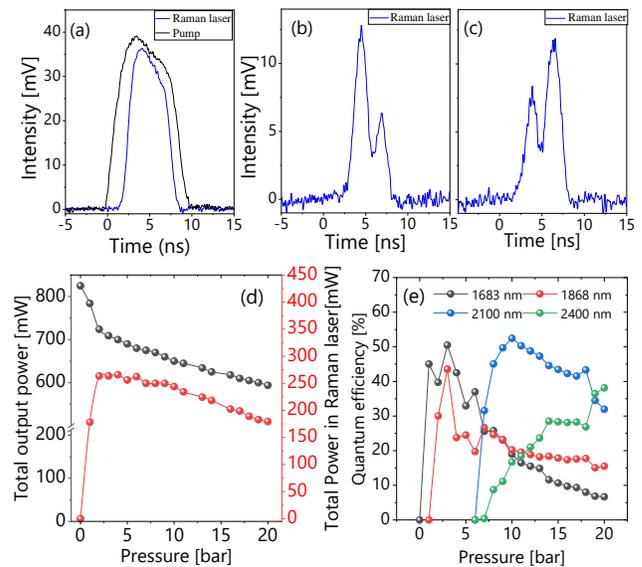

Figure 4: (a) Pulse profiles of first Raman Stokes at 1683 nm at ~1 bar (blue) and pump profile (black). (b) and (c) are other occasionally occurring profiles. (d) Total output power (black) and total power of all Stokes lines (red) versus pressure. (e)

Calculated quantum efficiency (QE) of each Raman Stokes line versus pressure.

Figure 4(d) shows the evolution of total power (includes all Stokes lines and residual pump) and total Raman laser power (includes all Stokes lines, but not the pump) as a function of gas pressure. It can be seen that the former continuously drops down as the gas pressure increases, indicating a constant energy transfer from the pump to the Raman laser Stokes lines. Meanwhile, the latter first quickly increases to a maximum value of 268 mW, but then drops slowly again due to the saturation of the Raman gain coefficient with increase in $H_2$ pressure [30]. This is mainly attributed to the low photon energy at longer wavelengths and the increased fiber propagation loss experienced by the higher-order Stokes lines at longer wavelengths. A similar drop in output power with increase in pressure was also noticed in [25], where they showed a drop of ~20% when the pressure was increased from 0.5 mbar to 6 bar. This drop (which is more pronounced at high pressures) was attributed to additional losses induced by strains in the fiber caused by pressure [25]. The power of each individual Stokes line is measured by extracting them with a series of long-pass filters. Figure 4(e) shows the quantum efficiency for each Raman Stokes line. The maximum quantum efficiencies at 1683 nm, 1868 nm, 2099 nm, and 2394 nm are 50.5%, 43.5%, 52.5%, and 38.1% respectively, corresponding to pulse energies of 18.5 µJ, 14.4 µJ, 14.1 µJ and 8.3 µJ, respectively. One can see that the pulse energy of each Stokes line can easily reach the micro-joule level. To the best of our knowledge, this is the first time to achieve micro-joule pulse energy over four cascaded Raman Stokes lines in such a broad spectral range. This benefits from not only the circularly polarized pump pulses with a high peak power of 13 kW and long pulse duration of 7 ns, but also from the relatively low transmission loss of the NARF at the pump wavelength and its small core diameter of 37 µm, which is necessary to enhance the light intensity and thereby increase the gain coefficient.

In conclusion, we have demonstrated a multi-wavelength fiber Raman laser spanning from 1.53 µm to 2.4 µm based on the cascaded rotational SRS effect in an $H_2$-filled NARF. The quantum efficiencies of all Stokes lines are higher than 38%, ensuring high pulse energies in the micro-joule level. This work is anticipated to have a series of applications such as supercontinuum generation through broadening of individual lines that could consequently create a broad spectrum, multi-species gas monitoring, photoacoustic spectroscopy, etc.

**Funding.** This work is supported by the Danmarks Frie Forskningsfond Hi-SPEC project (Grant No. 8022-00091B), Innovation Fund Denmark UVSUPER (Grant No. 8090-00060A), ECOMETA (Grant No. 6150-00030B) and US ARO (Grant No. W911NF-19-1-0426). Horizon 2020 research and innovation programme (Grant No. 722380).

**Disclosures.** The authors declare no conflicts of interest.